\documentclass[12pt,twoside]{article}
\usepackage{epsfig}
\usepackage{graphicx}
\usepackage{amssymb}
\usepackage{amsfonts,amsbsy}
\usepackage{amsmath}

\title{Plane-Wave Propagation in\\ Electromagnetic PQ Medium}
\author{I.V. Lindell} 
\date{Department of Radio Science and Engineering\\ Aalto University, School of Electrical Engineering\\Espoo, Finland\\ 
{\tt ismo.lindell@aalto.fi}\\ \today}
\def\e{\begin{equation}} 
\def\f{\end{equation}} 
\def\ea{\begin{eqnarray}} 
\def\fa{\end{eqnarray}} 

\def\##1{{\mbox{\textbf{#1}}}}
\def\%#1{{\mbox{\boldmath $#1$}}}
\def\=#1{{\overline{\overline{\mathsf #1}}}}

\def\SE{{\mathbb E}}
\def\SF{{\mathbb F}}

\def\*{^{\displaystyle*}}

\def\.{\cdot}
\def\x{\times}

\def\ra{\rightarrow}

\def\l#1{\label{eq:#1}}
\def\r#1{(\ref{eq:#1})}
\def\am{\left(\begin{array}{c}}
\def\amm{\left(\begin{array}{cc}}
\def\ammm{\left(\begin{array}{ccc}}
\def\ammmm{\left(\begin{array}{cccc}}
\def\a{\end{array}\right)}

\def\A{\alpha}
\def\B{\beta}

\def\De{\Delta}
\def\E{\epsilon}

\def\la{\lambda}

\def\M{\mu}

\def\t{\tau}
\def\z{\zeta}

\def\ve{\%\varepsilon}

\def\tr{{\rm tr }}

\def\W{\wedge}
\def\WW{\displaystyle{{}^\wedge}\llap{${}_\wedge$}}

\def\J{\rfloor}
\def\L{\lfloor}
\def\JJ{\rfloor\rfloor}
\def\LL{\lfloor\lfloor}

\begin{document}

\maketitle

\begin{abstract}
Two basic classes of electromagnetic media, recently defined and labeled as those of P media and Q media, are generalized to define the class of PQ media. Plane wave propagation in the general PQ medium is studied and the quartic dispersion equation is derived in analytic form applying four-dimensional dyadic formalism. The result is verified by considering various special cases of PQ media for which the dispersion equation is known to decompose to two quadratic equations or be identically satisfied (media with no dispersion equation). As a numerical example, the dispersion surface of a PQ medium with non-decomposable dispersion equation is considered. \end{abstract}

\section{Introduction}

The most general linear (bi-anisotropic) electromagnetic medium can be represented in terms of 36 scalar medium parameters in different representations. In engineering form applying Gibbsian vector fields and 3D dyadics it is typical to apply the form \cite{Kong,Methods}
\e \am \#D\\ \#B\a = \amm \=\E & \=\xi\\ \=\z & \=\M\a\.\am \#E\\ \#H\a. \l{DB}\f
Another form favored by the physicists is \cite{Hehl}
\e \am \#D\\ \#H\a = \amm \=\A & \=\E' \\ \=\M{}^{-1} & \=\B\a \.\am \#B\\ \#E\a. \l{DH}\f
A more compact form is obtained by applying the 4D formalism in terms of differential forms \cite{Deschamps,Difform} where the electromagnetic fields are characterized by two six-dimensional quantities, the two-forms $\%\Psi$ and $\%\Phi$,
\e \%\Psi = \#D - \#H\W\ve_4,\ \ \ \ \%\Phi=\#B + \#E\W\ve_4. \f
In a linear medium they are related by the medium bidyadic $\=M$ as
\e \%\Psi = \=M|\%\Phi. \f
The medium bidyadic can be expanded in terms of 3D medium dyadics as
\e \=M = \=\A + \=\E'\W\#e_4 + \ve_4\W\=\M{}^{-1} + \ve_4\W\=\B\W\#e_4. \f
For details in the present notation \cite{Difform} or \cite{MDEM} should be consulted.

It is not easy to get a grasp of the most general medium defined by the bidyadic $\=M$. This is why many classes of media with $\=M$ restricted by special forms involving less than 36 parameters, and bearing strange names, have been defined and studied. Because $\=M$ corresponds to a $6\x6$ matrix, some classes have been based on expressing the bidyadic $\=M$ in terms of a dyadic corresponding to a $4\x4$ matrix involving only 16 parameters. As the most obvious one, the class of P media \cite{P} has been defined in terms of a dyadic $\=P\in\SE_1\SF_1$ as the double-wedge square
\e \=M= \=P{}^{(2)T} = \frac{1}{2}\=P{}^T\WW\=P{}^T. \f
As another obvious example, the class of Q media \cite{416} has been defined through the modified medium bidyadic
\e \=M_m = \#e_N\L\=M = \#e_N\L\=Q{}^{(2)}\in\SE_2\SE_2, \f
in terms of a (quasi-) metric dyadic $\=Q\in\SE_1\SE_1$. Properties of both medium classes have been recently studied. For a plane wave, a field of the dependence $\exp(\%\nu|\#x)$ of the spacetime variable $\#x$, the wave one-form $\%\nu$ is normally restricted by a dispersion equation 
\e D(\%\nu)=0, \l{Dnu}\f
which is a quartic equation in general \cite{PIER05} and its form depends on the medium bidyadic $\=M$. For example, for the Q medium the dispersion equation is of the form
\e D(\%\nu)= \De_Q(\=Q||\%\nu\%\nu)^2=0, \l{DQ}\f
which means that, for a dyadic of full rank satisfying $\De_Q=\ve_N\ve_N||\=Q{}^{(4)}\not=0$, the quartic equation reduces to a quadratic equation. On the other hand, one can show that, for a P medium, the dispersion equation \r{Dnu} is actually an identity which is satisfied by any one-form $\%\nu$ \cite{P}. Also, if $\=Q$ is of rank lower than 4, we have $\De_Q=0$ and \r{DQ} is satisfied identically. Media of this kind have previously been called NDE media (media with no dispersion equation) \cite{NDE}. 

As simple generalizations of the classes of P and Q media we can define 
\e \=M = \=P{}^{(2)T} + \ve_N\L\#D\#C, \l{Pext}\f
and 
\e \=M_m = \=Q{}^{(2)} + \#A\#B, \l{Qext}\f
which have respectively been called extended (or generalized) P \cite{P} and Q \cite{421} media. Here, $\#A,\#B,\#C$ and $\#D$ are any bivectors. The number of medium parameters in both cases has been increased from 16 to 23.

Plane waves propagating in extended P and Q media have been previously studied and the dispersion equations have been shown to take the respective form \cite{P,421}
\ea D(\%\nu) &=& \De_P((\#D\L\=P{}^T)||\%\nu\%\nu)((\#C\L\=P{}^{-1T})||\%\nu\%\nu)=0, \l{DPext}\\
D(\%\nu) &=& \De_Q(\=Q||\%\nu\%\nu)(\=Q + \#A\#B\LL\=Q{}^{-1T})||\%\nu\%\nu=0. \l{DQext}\fa
Here we denote $\De_P=\tr\=P{}^{(4)}$. In both cases, for full-rank dyadics $\=P,\=Q$, the dispersion equations can be decomposed in two quadratic equations. Such media make two examples of what have been called decomposable media \cite{deco}. For dyadics $\=P$ and $\=Q$ of rank less than 3, the two media again fall in the class of NDE media with dispersion equations satisfied identically for any $\%\nu$ \cite{NDE}. For ranks equaling 3, some quadratic functions of $D(\%\nu)$ may decompose to products of linear functions of $\%\nu$.

\section{Class of PQ Media}

In the present study we make a further generalization of both P and Q media by considering medium bidyadics of the composite form
\e \=M = \=P{}^{(2)T} + \ve_N\L\=Q{}^{(2)}. \l{PQ}\f
Media defined by bidyadics of this form will be called PQ media. It is easy to see that medium dyadics of this form do not cover all possible linear media. In fact, the number of parameters must certainly be less than $2\x16=32$, which falls short of the 36 parameters of the most general medium. 

To find a 3D representation of the PQ medium bidyadic \r{PQ} the dyadics $\=P$ and $\=Q$ can be expanded in terms of spatial vectors $\#a_s,\#b_s,\#p_s$, a spatial one-form $\%\pi_s$ and spatial dyadics $\=Q_s,\=P_s$ as 
\ea \=Q &=& \=Q_s + \#e_4\#a_s + \#b_s\#e_4 + c\#e_4\#e_4, \\
\=P &=& \=P_s + \#e_4\%\pi_s + \#p_s\ve_4 + p\#e_4\ve_4. \fa
The spatial medium dyadics of the Q medium can be expressed as \cite{Difform}
\ea \=\A &=& \ve_{123}\L\=Q_s\W\#a_s, \l{AQ}\\
		\=\E' &=& \ve_{123}\L(c\=Q_s -\#b_s\#a_s), \\
		\=\M{}^{-1} &=& -\ve_{123}\L\=Q{}_s^{(2)}, \l{MQ}\\
		\=\B &=& \ve_{123}\L(\#b_s\W\=Q_s), \l{BQ}\fa		
while those of the P medium have the form \cite{P}
\ea \=\A &=& \=P{}_s^{(2)T}, \l{AP}\\
		\=\E' &=& -\%\pi_s\W\=P{}_s^T, \\
		\=\M{}^{-1}&=& -\=P{}_s^T\W\#p_s, \l{MP}\\
		\=\B &=& \%\pi_s\#p_s -p\=P{}_s^T. \l{PB}\fa
The medium dyadics of the PQ medium are obtained by summing up the corresponding expressions.	To write the medium equations in the Gibbsian form \r{DB} requires that the dyadic $\=\M$ exist. Denoting the Gibbsian dyadics by the subscript $()_g$, they are obtained by inserting the previous expansions in the expressions \cite{Difform}
\ea \=\E_g &=& \#e_{123}\L(\=\E'-\=\A|\=\M|\=\B) , \\	
\=\xi_g &=& \#e_{123}\L\=\A|\=\M, \\
\=\z_g &=& -\#e_{123}\L\=\M|\=\B, \\
\=\M_g &=&\#e_{123}\L\=\M. \fa
The dyadic denoted here by $\=\M$ stands for the inverse of the sum of the two $\=\M{}^{-1}$ dyadics of \r{MQ} and \r{MP}. The full analytic form of the Gibbsian medium dyadics of the PQ medium would have quite an extensive form.

\section{Plane Wave in PQ medium}

The main task of this study is to find properties of a plane wave propagating in the general PQ medium. Expressing the field two-form of a plane wave in terms of a potential one-form as $\%\Phi=\%\nu\W\%\phi$, the potential satisfies
\ea \%\nu\W\%\Psi &=& \%\nu\W\=M|(\%\nu\W\%\phi) =0\nonumber\\
&=& \%\nu\W(\=P{}^{(2)T} + \ve_N\L\=Q{}^{(2)})|(\%\nu\W\%\phi) \nonumber\\
&=& \%\nu\W(\=P{}^T|\%\nu)\W(\=P{}^T|\%\phi) + \ve_N\L(\=Q{}^{(2)}\LL\%\nu\%\nu)|\%\phi . \fa
Operating this by $\#e_N\L$ yields the equation
\ea \#e_N\L(\%\nu\W\%\Psi) &=& \#e_N\L(\%\nu\W(\%\nu|\=P)\W\=P{}^T)|\%\phi + (\=Q{}^{(2)}\LL\%\nu\%\nu)|\%\phi \nonumber\\
&=& (\#F\L\=P{}^T +\=Q{}^{(2)}\LL\%\nu\%\nu)|\%\phi=0. \l{eq}\fa
Here we have introduced the bivector
\e \#F = \#F(\%\nu)= \#e_N\L(\%\nu\W(\%\nu|\=P)), \f
which is simple since it satisfies \cite{MDEM}
\e  \#F\W\#F=0,\ \ \ \ (\#F\L\=I{}^T)^{(2)}=\#F\#F,\ \ \ \ (\#F\L\=I{}^T)^{(3)}=0, \f
and it can be expressed  in the form $\#F=\#a\W\#b$ in terms of two vectors. The equation \r{eq} for the potential, 
\e \=D(\%\nu)|\%\phi=0, \l{Dnuphi}\f
is defined by the dispersion dyadic
\e \=D(\%\nu) = \#F\L\=P{}^T +\=Q{}^{(2)}\LL\%\nu\%\nu. \f
Because of \r{Dnuphi} and $\=D(\%\nu)|\%\nu=0$, the rank of $\=D(\%\nu)$ must be less than three, provided $\%\phi$ and $\%\nu$ are linearly independent, i.e., when $\%\Phi=\%\nu\W\%\phi\not=0$, which is assumed here. As a consequence, $\%\nu$ is restricted by the dyadic dispersion equation \cite{MDEM}
\ea \=D{}^{(3)}(\%\nu) &=& (\#F\L\=P{}^T +\=Q{}^{(2)}\LL\%\nu\%\nu)^{(3)} \nonumber\\
&=& (\#F\L\=I{}^T)^{(3)}|\=P{}^{(3)T} + ((\#F\L\=I{}^T)^{(2)}|\=P{}^{(2)T})\WW(\=Q{}^{(2)}\LL\%\nu\%\nu) \nonumber\\
&&+ (\#F\L\=P{}^T)\WW(\=Q{}^{(2)}\LL\%\nu\%\nu)^{(2)} + (\=Q{}^{(2)}\LL\%\nu\%\nu)^{(3)} \nonumber\\
&=&  (\#F\#F|\=P{}^{(2)T})\WW(\=Q{}^{(2)}\LL\%\nu\%\nu) + (\#F\L\=P{}^T)\WW(\=Q{}^{(2)}\LL\%\nu\%\nu)^{(2)} + (\=Q{}^{(2)}\LL\%\nu\%\nu)^{(3)} \nonumber\\
&=& 0. \l{Disp3}\fa
Applying the expansion rules \cite{MDEM}
\ea \=Q{}^{(2)}\LL\%\nu\%\nu &=& (\=Q||\%\nu\%\nu)\=Q - (\=Q|\%\nu)(\%\nu|\=Q) \nonumber\\
(\=Q{}^{(2)}\LL\%\nu\%\nu)^{(2)} &=& (\=Q||\%\nu\%\nu)((\=Q||\%\nu\%\nu)\=Q{}^{(2)} - \=Q\WW(\=Q|\%\nu)(\%\nu|\=Q))\nonumber\\
&=& (\=Q||\%\nu\%\nu)(\=Q{}^{(3)}\LL\%\nu\%\nu) \\
(\=Q{}^{(2)}\LL\%\nu\%\nu)^{(3)} &=& (\=Q||\%\nu\%\nu)^2((\=Q||\%\nu\%\nu)\=Q{}^{(3)} - \=Q{}^{(2)}\WW(\=Q|\%\nu)(\%\nu|\=Q) \nonumber\\
&=& (\=Q||\%\nu\%\nu)^2(\=Q{}^{(4)}\LL\%\nu\%\nu)=  \De_Q(\=Q||\%\nu\%\nu)^2(\#e_N\#e_N\LL\%\nu\%\nu), \fa
with $\De_Q=\ve_4\ve_4||\=Q{}^{(4)}$, the dispersion equation \r{Disp3} can be written as
\e \=D{}^{(3)}(\%\nu) = \=C_1(\%\nu) + \=C_2(\%\nu) + \=C_3(\%\nu) =0, \l{D3nu}\f
where we denote
\ea \=C_1(\%\nu) &=& (\#F\#F|\=P{}^{(2)T})\WW(\=Q{}^{(2)}\LL\%\nu\%\nu) \\
\=C_2(\%\nu) &=&  (\=Q||\%\nu\%\nu)(\#F\L\=P{}^T)\WW(\=Q{}^{(3)}\LL\%\nu\%\nu) \\
\=C_3(\%\nu) &=& \De_Q(\=Q||\%\nu\%\nu)^2(\#e_N\#e_N\LL\%\nu\%\nu).\fa
Now one can show that \r{D3nu} is equivalent to a scalar dispersion equation \r{Dnu}. For that we expand the three dyadics $\=C_i(\%\nu)$ as follows. 

\begin{itemize}

\item For the dyadic $\=C_1(\%\nu)$ we apply the identity
\ea \#A_i\W(\#B_i\L\%\A) &=& (\#A_i\W\#B_i)\L\%\A - \#B_i\W(\#A_i\L\%\A)\nonumber\\
&=& \ve_N|(\#A_i\W\#B_i)(\#e_N\L\%\A) - \#B_i\W(\#A_i\L\%\A), \fa
valid for any bivectors $\#A_i,\#B_i$ and one-form $\%\A$. Assuming that $\=A_i$ and $\%\A$ satisfy $\=A_i\L\%\A=0$, we can construct the dyadic rule
\e \#A_1\#A_2\WW(\#B_1\#B_2\LL\%\A\%\A) = \ve_N\ve_N||(\#A_1\#A_2\WW\#B_1\#B_2)(\#e_N\#e_N\LL\%\A\%\A). \f
Because of $\#F\L\%\nu=0$ and $(\#F|\=P{}^{(2)T})\L\%\nu = (\#F\L(\%\nu|\=P))\L\=P{}^T=0$, we can set $\#A_1=\#F$, $\#A_2=\#F|\=P{}^{(2)T}$ and $\%\A=\%\nu$. Since the rule is linear in the dyadic $\#B_1\#B_2$, we can set $\#B_1\#B_2=\=Q{}^{(2)}$ and apply the rule as  
\e \=C_1(\%\nu) = (\#F\#F|\=P{}^{(2)T})\WW(\=Q{}^{(2)}\LL\%\nu\%\nu) = D_1(\%\nu)(\#e_N\#e_N\LL\%\nu\%\nu), \f
with
\ea D_1(\%\nu) &=& \ve_N\ve_N||((\#F\#F|\=P{}^{(2)T})\WW(\=Q{}^{(2)})) =(\ve_N\ve_N\LL\=Q{}^{(2)})||(\#F\#F|\=P{}^{(2)T}) \nonumber\\
&=&\De_Q \=Q{}^{(-2)T}||(\#F\#F|\=P{}^{(2)T}) = \De_Q\#F\#F||(\=P{}^T|\=Q{}^{-1})^{(2)} . \l{D1}\fa 
In the last expression we have assumed that $\=Q$ is of full rank, $\De_Q\not=0$.  '
\item To expand the dyadic $\=C_2(\%\nu)$ we proceed as
\ea \=C_2(\%\nu) &=&  (\=Q||\%\nu\%\nu)(\=Q{}^{(3)}\LL\%\nu\%\nu)\WW(\#F\L\=P{}^T) \nonumber\\
&=&  \De_Q(\=Q||\%\nu\%\nu)((\#e_N\#e_N\LL(\=Q{}^{-1T}\WW\%\nu\%\nu))\WW(\#F\L\=P{}^T) \nonumber\\
&=&  \De_Q(\=Q||\%\nu\%\nu)\#e_N\#e_N\LL((\=Q{}^{-1T}\WW\%\nu\%\nu)\LL(\#F\L\=P{}^T)) \nonumber\\
&=&  \De_Q(\=Q||\%\nu\%\nu)(\#e_N\#e_N\LL\%\nu\%\nu)(\=Q{}^{-1T}||(\#F\L\=P{}^T))\nonumber\\
&=&  \De_Q(\=Q||\%\nu\%\nu)(\#e_N\#e_N\LL\%\nu\%\nu)(\=Q{}^{-1T}||(\#F\L\=P{}^T))\nonumber\\
&=&  D_2(\%\nu)(\#e_N\#e_N\LL\%\nu\%\nu), \fa
with
\e D_2(\%\nu) = \De_Q(\=Q||\%\nu\%\nu)\tr(\#F\L\=P{}^T|\=Q{}^{-1}). \f

\item Finally, we have
\e \=C_3(\%\nu) = D_3(\%\nu)(\ve_N\ve_N\LL\%\nu\%\nu), \f
with
\e D_3(\%\nu) = \De_Q (\=Q||\%\nu\%\nu)^2. \l{D3}\f
\end{itemize}

Because each of the dyadics $\=C_i(\%\nu)$ is a scalar multiple of $\#e_N\#e_N\LL\%\nu\%\nu$, the dyadic dispersion equation \r{D3nu} equals the scalar dispersion equation \r{Dnu} as
\ea D(\%\nu) &=& D_1(\%\nu) + D_2(\%\nu) + D_3(\%\nu) \nonumber\\
&=& \De_Q\#F\#F||(\=P{}^T|\=Q{}^{-1})^{(2)} + \De_Q(\=Q||\%\nu\%\nu)\tr(\#F\L\=P{}^T|\=Q{}^{-1}) \nonumber\\
&&+\De_Q (\=Q||\%\nu\%\nu)^2 =0. \l{Disp}\fa
Substituting $\#F=\#e_N\L(\%\nu\W(\%\nu|\=P))$, the quartic form of the dispersion  equation can be . This is the main result of the present paper. 

For $\De_Q\ra0$ we must replace  
\e \De_Q \=Q{}^{(-2)} \ra \ve_N\ve_N\LL\=Q{}^{(2)T},\ \ \ \ \De_Q\=Q{}^{-1} \ra \ve_N\ve_N\LL\=Q{}^{(3)T}, \l{DeQ}\f
in the expression \r{Disp}.

\section{Special Cases}

Let us consider the expression \r{Disp} for a few special cases of the PQ medium for which we know the dispersion equation.

\begin{enumerate}
\item For the pure P medium case $\=Q=0$, after inserting \r{DeQ} we obtain the identity $D(\%\nu)=0$ for all $\%\nu$. This proves that a pure P medium belongs to the class of NDE media \cite{NDE}.

\item For the pure Q medium with $\=P=0$ \r{Disp} is reduced to the known quadratic dispersion equation \r{DQ}. 

 \item The case $\=P = \A\=I$ corresponds to a Q-axion medium. Since we now have $\#F=\A\#e_N\L(\%\nu\W\%\nu)=0$, only the last term of the expression in \r{Disp} survives. This, again, yields the dispersion equation of the Q medium. In fact, it is well known that adding an axion term $\A\=I{}^{(2)T}$ to the medium bidyadic does not change the dispersion equation. Here one should note that the P-axion medium $\=M=\=P{}^{(2)T}+ \A\=I{}^{(2)T}$ with $\A\not=0$ is not a special case of the PQ medium.

\item Choosing $\=Q=\#a_1\#b_1+ \#a_2\#b_2$ we obtain 
\e\=M = \=P{}^{(2)T} + \ve_N\L(\#a_1\W\#a_2)(\#b_1\W\#b_2), \f 
which yields a special case of the extended P medium \r{Pext} with $\#D\#C=(\#a_1\W\#a_2)(\#b_1\W\#b_2)$. Applying \r{DeQ} for  $\De_Q\ra0$ in \r{Disp} yields the dispersion equation whose form can be expanded as
\ea D(\%\nu) &=& \#F|(\=P{}^{(2)T}|\=Q{}^{(2)T})|\#F \nonumber\\
&=& \#F|\=P{}^{(2)T}|(\ve_N\L(\#b_1\W\#b_2))((\#a_1\W\#a_2)\J\ve_N)|\#F \nonumber\\
&=& (\%\nu\W(\%\nu|\=P))\J\#e_N|\=P{}^{(2)T}|\ve_N\L(\#b_1\W\#b_2))((\#a_1\W\#a_2)|(\%\nu\W(\%\nu|\=P)) \nonumber\\
&=& \De_P(\%\nu\W(\%\nu|\=P)|\=P{}^{(-2)}|(\#b_1\W\#b_2))((\#a_1\W\#a_2)|(\%\nu\W(\%\nu|\=P)) \nonumber\\
&=& \De_P((\%\nu|\=P{}^{-1})\W\%\nu)|(\#b_1\W\#b_2))((\#a_1\W\#a_2)|(\%\nu\W(\%\nu|\=P)) =0.\fa
Assuming $\De_P=\tr\=P{}^{(4)}\not=0$ the equation is split in two
quadratic equations as
\ea \%\nu|(\=P\W\%\nu)|(\#a_1\W\#a_2) &=& \%\nu|(\=P\J(\#a_1\W\#a_2))|\%\nu =0, \\
\%\nu|(\=P{}^{-1}\W\%\nu)|(\#b_1\W\#b_2) &=& \%\nu|(\=P{}^{-1}\J(\#b_1\W\#b_2))|\%\nu=0 , \l{nuP-1}\fa
which coincide with \r{DPext} for $\#D\#C=(\#a_1\W\#a_2)(\#b_1\W\#b_2)$. In the case $\De_P=0$ we must replace $\=P{}^{-1}$ by $\#e_N\ve_N\LL\=P{}^{(3)T}$ in \r{nuP-1}.

\item Choosing $\=P = \#a_1\%\A_1+ \#a_2\%\A_2$ we have $\=P{}^{(3)}=0$ and
\e\=M= \ve_N\L\=Q{}^{(2)T} +(\%\A_1\W\%\A_2)(\#a_1\W\#a_2), \f
which corresponds to a special case of the extended Q medium \r{Qext} with $\#A\#B=\#e_N\L(\%\A_1\W\%\A_2)(\#a_1\W\#a_2)$. 
Expanding 
\e \#F|\=P{}^{(2)T} = \#e_N\L(\%\nu\W(\=P{}^T|\%\nu))|\=P{}^{(2)T}= \#e_N|(\%\nu\W(\=P{}^{(3)T}\L\%\nu)) =0, \f
the first term of \r{Disp} obviously vanishes. Expanding further
\ea \#F\L\=P{}^T &=& \#e_N\L(\%\nu\W(\=P{}^T|\%\nu)\W\=P{}^T) =\#e_N\L(\%\nu\W(\=P{}^{(2)T}\L\%\nu))\nonumber\\
&=& -\%\nu\J(\#e_N\L\=P{}^{(2)T})\L\%\nu = (\#e_N\L\=P{}^{(2)T})\LL\%\nu\%\nu, \fa
we have
\e \tr(\#F\L\=P{}^T|\=Q{}^{-1}) = (\#e_N\L\=P{}^{(2)T})\LL\%\nu\%\nu)||\=Q{}^{-1T} = ((\#e_N\L\=P{}^{(2)T})\LL\=Q{}^{-1T})||\%\nu\%\nu, \f
whence the dispersion equation \r{Disp} can be written as
\e D(\%\nu) = \De_Q(\=Q||\%\nu\%\nu)(\=Q + (\#e_N\L\=P{}^{(2)T})\LL\=Q{}^{-1T})||\%\nu\%\nu=0, \f
which coincides with \r{DQext}. 

\item Finally, let us assume that $\=Q$ is an antisymmetric dyadic, which can be expressed in terms of some bivector $\#A$ in the form
\e \=Q = \#A\L\=I{}^T. \f
This implies $\=Q||\%\nu\%\nu=0$, whence the dispersion equation \r{Disp} reduces to
\e D(\%\nu) = \De_Q\#F|\=P{}^{(2)T}|\=Q{}^{(-2)}|\#F=0. \l{Disp1} \f
Applying the expansion \cite{MDEM}
\e (\#A\L\=I{}^T)^{(2)} = \#A\#A +\A\#e_N\L\=I{}^{(2)T}, \ \ \ \ \A=-\frac{1}{2}\ve_N|(\#A\W\#A), \f
the PQ medium bidyadic has the form of an extended P-axion medium bidyadic \cite{deco}
\e \=M = \=P{}^{(2)T} + \ve_N\L\#A\#A + \A\=I{}^{(2)T}. \f
Applying the rule \cite{MDEM} 
\e \=Q{}^{(-2)} = \frac{1}{\De_Q}\ve_N\ve_N\LL\=Q{}^{(2)T} = \frac{1}{\De_Q}(\ve_N\ve_N\LL\#A\#A + \A\ve_N\L\=I{}^{(2)}), \f
one can show that its last term has no effect on the dispersion equation. In fact, inserting
\ea \#F|\=P{}^{(2)T} &=& \#e_N|(\%\nu\W(\=P{}^T|\%\nu)\W\=P{}^{(2)T}) =\%\nu|(\#e_N\L\=P{}^{(3)T})\L\%\nu \nonumber\\
&=&  \De_P\%\nu|(\=P{}^{-1}\J\#e_N)\L\%\nu, \fa
in \r{Disp1} we have
\ea \#F|\=P{}^{(2)T}|(\ve_N\L\=I{}^{(2)})|\#F &=& \De_P(\%\nu|(\=P{}^{-1}\J\#e_N)\L\%\nu)|(\ve_N\L(\#e_N\L(\%\nu\W(\=P{}^T|\%\nu))) \nonumber\\
&=& \De_P(\%\nu|(\=P{}^{-1}\J\#e_N)|(\%\nu\W\%\nu\W(\=P{}^T|\%\nu))=0. \fa
Thus, \r{Disp1} is reduced to
\ea D(\%\nu) &=& \De_Q\#F|\=P{}^{(2)T}|\=Q{}^{(-2)}|\#F \nonumber\\
&=&\De_P\%\nu|(\=P{}^{-1}\J\#e_N)\L\%\nu|(\ve_N\ve_N\LL\#A\#A)|(\#e_N\L(\%\nu\W(\=P{}^T|\%\nu)) \nonumber\\
&=&\De_P\%\nu|(\=P{}^{-1}\J\#e_N)\L\%\nu|(\ve_N\L\#A)(\#A|(\%\nu\W(\=P{}^T|\%\nu)))\nonumber\\
&=&\De_P((\%\nu|\=P{}^{-1}\W\%\nu)|\#A)(\#A|(\%\nu\W(\=P{}^T|\%\nu)))\nonumber\\
&=&\De_P((\=P{}^{-1}\J\#A)||\%\nu\%\nu)((\=P\J\#A)||\%\nu\%\nu)=0. \fa
Since the axion component does not affect the dispersion equation, the result coincides with that of the extended P medium \r{DPext} for the special case  $\#C=\#D=\#A$. 

\end{enumerate}

\section{Example}

As a numerical example of a PQ medium let us consider a special one by restricting the dyadic $\=P$ as
\e \=P = P(\=I + \=P_o),\ \ \ \ \=P_o=\#a\%\A,\ \ \ \ \#a|\%\A=0,\f
and assuming that $\=Q$ be a symmetric dyadic,
\e \=Q=\=S,\ \ \ \  \=S{}^T=\=S.\f
In the corresponding medium bidyadic
\e \=M = P^2\=I{}^{(2)T} + P^2\%\A\#a\WW\=I{}^T + \ve_N\L\=S{}^{(2)} \l{MS2}\f
the three terms can actually be recognized as components of the Hehl-Obukhov decomposition, respectively called as the axion, skewon and principal components \cite{Hehl,MDEM}.  In fact, the axion part is a multiple of the unit bidyadic $\=I{}^{(2)T}$ while the skewon and principal parts are trace free. Also, any skewon bidyadic is known to be of the form $(\=B_o\WW\=I)^T$ with a skewon-free dyadic $\=B_o$ while any bidyadic $\=C\in\SF_2\SE_2$ satisfying $\=C\LL\=I=0$ is a principal bidyadic. Actually, one can show that $(\ve_N\L\=S{}^{(2)})\LL\=I$ vanishes for any symmetric dyadic $\=S$, \cite{MDEM}. 

To find the dispersion equation for the present PQ medium from \r{Disp}, let us first expand
\ea \#F|\=P{}^{(2)T} &=& \#e_N|(\%\nu\W(\=P{}^{(3)T}\L\%\nu)) \nonumber\\
&=& P^3\#e_N|(\%\nu\W(\=I{}^{(3)T}+ \%\A\#a\WW\=I{}^{(2)T}))\L\%\nu \nonumber\\
&=& P^3\#e_N\L(\%\nu\W\%\nu\W\=I{}^{(2)T}) + P^3 \#e_N|(\%\nu\W(\%\A(\#a|\%\nu)\W\=I{}^{(2)T}\nonumber\\
&& + \%\nu\W(\%\A\#a)\W\=I{}^T) \nonumber\\
&=& P^3(\#a|\%\nu)\#e_N\L(\%\nu\W\%\A), \l{FP2T}\\
\#F\L\=P{}^T&=& \#e_N\L(\%\nu\W(\=P{}^{(2)T}))\L\%\nu) \nonumber\\
&=& P^2\#e_N\L(\%\nu\W(\=I{}^{(2)T}+\%\A\#a\WW\=I{}^T)\L\%\nu) \nonumber\\
&=& P^2\#e_N\L(\%\nu\W\%\nu\W\=I{}^T+\%\nu\W(\%\A(\#a|\%\nu)\W\=I{}^T+ \%\nu\W\%\A\#a)) \nonumber\\
&=& P^2(\#a|\%\nu)(\#e_N\L(\%\nu\W\%\A))\L\=I{}^T. \fa
Since the last expression is an antisymmetric dyadic and $\De_S\=S{}^{-1}= \ve_N\ve_N\LL\=S{}^{(3)}$ is a symmetric dyadic or zero, we have
\e \De_Q\tr(\#F\L\=P{}^T|\=Q{}^{-1}) = \De_S(\#F\L\=P{}^T)||\=S{}^{-1}=0. \f
Thus, \r{Disp} becomes
\e D(\%\nu) = \De_S\#F|(\=P{}^T|\=S{}^{-1})^{(2)}|\#F +\De_S (\=S||\%\nu\%\nu)^2. \f
Finally, applying \r{FP2T}, we can expand
\ea \De_S\#F|(\=P{}^T|\=S{}^{-1})^{(2)}|\#F&=& P^4(\#a|\%\nu)(\#e_N\L(\%\nu\W\%\A))|\=S{}^{(2)}|(\#e_N\L(\%\nu\W(\=I{}^T+\%\A\#a)|\%\nu))) \nonumber\\
&=& P^4(\#a|\%\nu)^2(\%\nu\W\%\A)|\=S{}^{(2)}|(\%\nu\W\%\A)) \nonumber\\
&=& P^4(\#a|\%\nu)^2(\%\A\%\A\JJ\=S{}^{(2)})||\%\nu\%\nu, \nonumber\fa
whence the dispersion equation for the special PQ medium \r{MS2} has the form
\ea D(\%\nu) &=& P^4(\#a|\%\nu)^2(\%\A\%\A\JJ\=S{}^{(2)})||\%\nu\%\nu+\De_S (\=S||\%\nu\%\nu)^2 \nonumber\\
&=& P^4(\#a|\%\nu)^2(\%\A\%\A||\=S)(\=S||\%\nu\%\nu) +P^4(\#a|\%\nu)^2(\%\A|\=S|\%\nu)^2\nonumber\\
&&+\De_S (\=S||\%\nu\%\nu)^2=0. \l{DispSpec}\fa
Unlike for all the special cases considered above, the quartic dispersion equation corresponding to the medium defined by \r{MS2} does not necessarily decompose in two quadratic equations. In the special case when the dyadic $\=S$ satisfies $\=S|\%\A=\la\#a$, whence we have $\=S||\%\A\%\A=0$, the dispersion equation is decomposable.

Setting $\=S=0$ in \r{DispSpec} we obtain $D(\%\nu)=0$ for any $\%\nu$, a property shared by all skewon-axion media. For $\#a\%\A=0$ we are left with $\De_S(\=S||\%\nu\%\nu)^2=0$, valid for the Q-axion medium. 

To be able to work on this example numerically, let us choose
\e \=S = S\=G_s - s\#e_4\#e_4, \f
where $\=G_s$ denotes the metric dyadic
\e \=G_s = \#e_1\#e_1 + \#e_2\#e_2+ \#e_3\#e_3. \f
Thus, we have
\e \De_S = \tr\=S{}^{(4)} = -S^3s. \f
Let us further choose
\e \%\A=\ve_2,\ \ \ \ \#a=\#e_3. \f
Inserting these in \r{DispSpec}, the dispersion equation becomes
\ea D(\%\nu) &=& 
P^4\nu_3^2S(S(\nu_1^2+\nu_2^2+\nu_3^2)-sk^2) +P^4\nu_3^2S^2\nu_2^2  \nonumber\\
&&-S^3s (S(\nu_1^2+\nu_2^2+\nu_3^2)-sk^2)^2=0. \l{DispSpec1}\fa
For a given value of $k$, this corresponds to a quartic dispersion surface in the three-dimensional space spanned by $\nu_i$. 

Let us assume that $S,s$ and $P$ are real and positive quantities and simplify the expressions by denoting 
\e x_i=\nu_i\sqrt{S},\ \ \ K = k\sqrt{s},\ \ \  \t=\frac{P^4}{S^3s}, \f
with $0\leq\t\leq1$. The dispersion equation \r{DispSpec1} then takes the form
\e  (x_1^2+x_2^2+(1-\t)x_3^2-K^2)(x_1^2+x_2^2+x_3^2-K^2) -\t x_3^2x_2^2 =0. \l{DispSpec2}\f

An idea of the dispersion surface can be obtained by considering the three main sections.
\begin{itemize}
\item Assuming $x_3=0$,  \r{DispSpec2} leads to
\e  x_1^2+x_2^2-K^2=0,\f
which corresponds to a circle of unit radius. 
\item Assuming $x_2=0$,  \r{DispSpec2} yields
\e (x_1^2+(1-\t)x_3^2-K^2)(x_1^2+x_3^2-K^2)=0. \f
This splits in two separate curves, one of which is a circle of unit radius 
\e x_1^2+x_3^2-K^2=0, \f
and, the other one, a quadratic curve
\e x_1^2+(1-\t)x_3^2-K^2=0. \f 
For $\t<1$ the latter defines an ellipse with axial ratio $\sqrt{1-\t}$. For $\t>1$ the curve is a hyperbola.
\item Finally, assuming $x_1=0$,  \r{DispSpec2} yields 
\e (x_2^2+(1-\t)x_3^2-K^2)(x_2^2+x_3^2-K^2) -\t x_3^2x_2^2 =0. \l{Dx1}\f
This corresponds to a curve of the fourth order, the form of which depends on the parameter $\t$. For $\t<1$ the curve is closed and for $\t\geq1$ it is open. For $\t\ra0$ the curve approaches a circle of unit radius, in which case the PQ medium approaches a Q medium. 
\end{itemize}

\begin{figure}
\includegraphics[width=\textwidth]{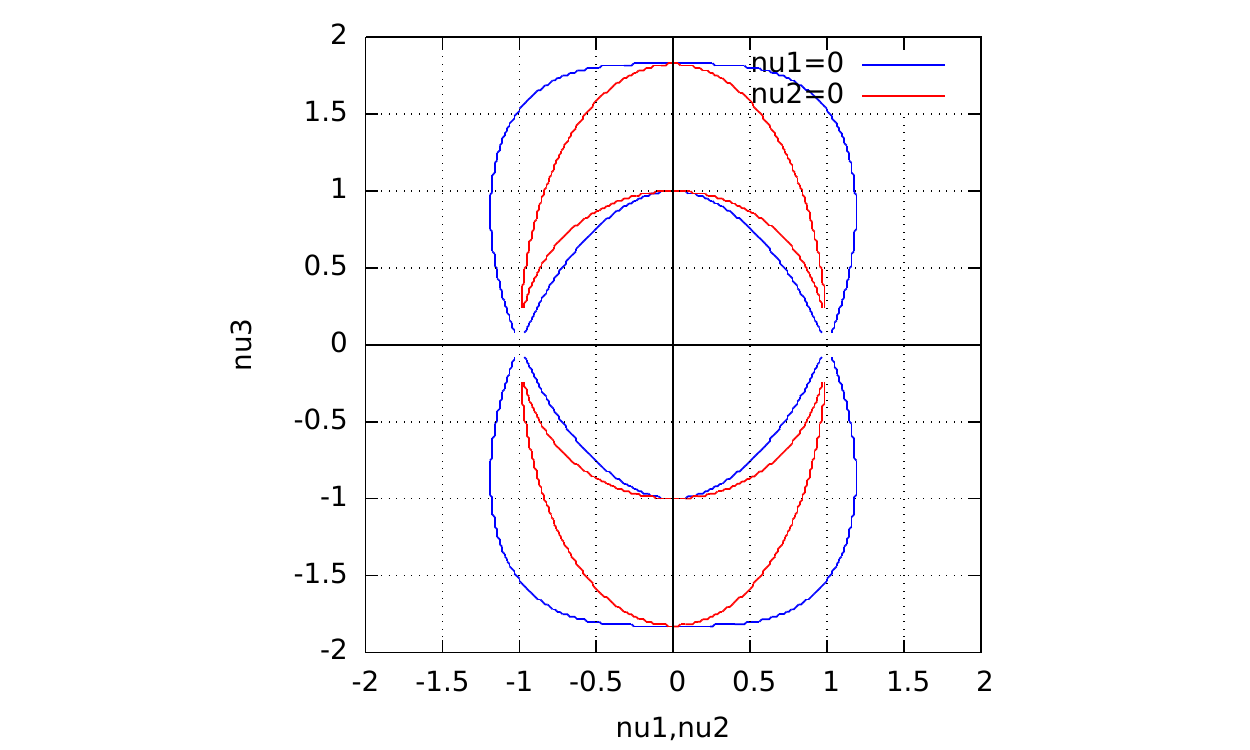}
\caption{Two cross sections of the quartic dispersion surface corresponding to $\nu_1=\#e_1|\%\nu=0$ and $\nu_2=\#e_2|\%\nu=0$. For $\nu_2=0$ the cross-section reduces to a circle and an ellipse. For $\nu_3=0$ the dispersion surface reduces to a single circle of unit radius. Here the parameter value $\t=P^4/S^3s=0.7$ has been assumed.}
\label{fig:EM26a08}
\end{figure}

The cross sections $x_1=0$ and $x_2=0$ are depicted for the parameter value $\t=0.7$ in Fig.\ \ref{fig:EM26a08} in terms of normalized axis parameters ${\rm nu}i=x_i/K$. It is seen that, for this particular PQ medium, there is no birefringence for waves whose wave one-form satisfies $\#e_3|\%\nu=0$.

\section{Conclusion}

A novel class of electromagnetic media, called that of PQ media, was introduced as a generalization of the previously studied classes of P media and Q media. Plane-wave propagating in the general PQ medium was studied and the quartic dispersion equation was derived in analytic form. The equation was verified for six special cases of PQ media for which the analytic form has been found from previous studies. In all of these special cases the quartic equation either reduces to two quadratic equations or becomes an identity. As an example of a medium yielding a more general quartic dispersion equation, another special case of the PQ medium was considered.

\section{Acknowledgment} Discussion with Dr. Alberto Favaro on the topic of this paper is acknowledged.

\end{document}